\documentclass[epsf,usegraphicx]{mn2e}

\usepackage{xspace}

\title[Stellar Pulsators at large distances] {New short period
stellar pulsators at large Galactocentric distances}

\author[]
{Gavin Ramsay$^{1}$, Ralf Napiwotzki$^{2}$, Thomas Barclay$^{1,3}$,
Pasi Hakala$^{4}$, Stephen Potter$^{5}$,  \and Mark Cropper$^{3}$ \\
$^{1}$Armagh Observatory, College Hill, Armagh, BT61 9DG\\
$^{2}$Centre for Astrophysics Research, STRI, University of Hertfordshire, 
Hatfield, AL10 9AB\\
$^{3}$Mullard Space Science Laboratory, University College London,
Holmbury St. Mary, Dorking, Surrey, RH5 6NT\\
$^{4}$Finnish Centre for Astronomy with ESO, University of Turku,
V\"{a}is\"{a}l \"{a}ntie 20, FI-21500 PIIKKI\"{O}, Finland\\ 
$^{5}$South African Astronomical Observatory, P.O. Box 9, Observatory
7935, Cape Town, South Africa\\}

\date{Accepted 2011 June 16.  Received 2011 June 15; in original form 2011 
April 25}

\begin{document}
\outer\def\gtae {$\buildrel {\lower3pt\hbox{$>$}} \over 
{\lower2pt\hbox{$\sim$}} $}
\outer\def\ltae {$\buildrel {\lower3pt\hbox{$<$}} \over 
{\lower2pt\hbox{$\sim$}} $}
\newcommand{\ergscm} {ergs s$^{-1}$ cm$^{-2}$}
\newcommand{\ergss} {ergs s$^{-1}$}
\newcommand{\ergsd} {ergs s$^{-1}$ $d^{2}_{100}$}
\newcommand{\pcmsq} {cm$^{-2}$}
\newcommand{\ros} {\sl ROSAT}
\newcommand{\chan} {\sl Chandra}
\newcommand{\xmm} {\sl XMM-Newton}
\def\rchi{{${\chi}_{\nu}^{2}$}}
\newcommand{\Msun} {$M_{\odot}$}
\newcommand{\Mwd} {$M_{wd}$}
\def\Mdot{\hbox{$\dot M$}}
\def\mdot{\hbox{$\dot m$}}
\newcommand{\teff}{\ensuremath{T_{\mathrm{eff}}}\xspace}
\newcommand{\porb}{\ensuremath{P_{\mathrm{orb}}}\xspace}

\maketitle

\begin{abstract}

We report the discovery of 31 blue, short period, pulsators made using
data taken as part of the Rapid Temporal Survey (RATS). We find they
have periods between 51--83 mins and full-amplitudes between
0.05--0.65 mag. Using the period-luminosity relationship for short
period pulsating stars we determine their distance. Assuming they are
pulsating in either the fundamental or first over-tone radial mode the
majority are located at a distance greater than 3kpc, with several
being more than 20 kpc distant. Most stars are at least 1 kpc from the
Galactic plane, with three being more than 10 kpc. One is located in
the direction of the Galactic anti-center and has Galactocentric
distance of $\sim$30 kpc and is $\sim$20 kpc below the plane: they are
therefore potential tracers of Galactic structure. We have obtained
low-resolution spectra for a small number our targets and find they
have temperatures between 7200--7900K and a metal content less than
Solar. The colours of the pulsators and the spectral fits to those
stars for which we have spectra indicate that they are either SX Phe
or $\delta$ Scuti stars.  We estimate the number of SX Phe stars in
our Galaxy and find significantly fewer per unit mass than reported in
massive globular clusters or dwarf spheroidal galaxies.

\end{abstract}


\section{Introduction}

Pulsations have been detected from stars on a wide range of
timescales, from several tens of seconds in the case of white dwarfs
to several years in the case of red-giant stars. These pulsations
manifest themselves through a periodic variation of the stellar
brightness. Pulsating stars can also be found over a wide range of
parameter space (temperature, luminosity) in the HR diagram (eg
Jeffery 2008). A detailed study of the photometric variability of
individual systems can give insight to the physical conditions deep
inside the star (eg Kurtz 2004).

In recent years many photometric surveys have been undertaken, leading
to a corresponding increase in the number of known stellar
pulsators. Factors such as cadence, depth, sky coverage and duration
makes any individual survey more (or less) likely to discover specific
types of stellar pulsator. The Rapid Temporal Survey (RATS) is a deep,
high-cadence photometric survey covering nearly 40 square degrees
which took place between 2003 and 2010 (Ramsay \& Hakala 2005, Barclay
et al 2011).  This strategy allows us to detect sources which vary in
their intensity on a timescale of a few minutes to several hours.

In our first set of wide-field camera data taken in 2003, we
identified a small number of blue stars which pulsate on a period
between 40--70 mins. An analysis of their optical spectra indicated
they were SX Phe stars or $\delta$ Sct stars (Ramsay et al 2006).  SX
Phe stars are old, metal-poor stars which are likely to be halo
objects (see Nemec \& Mateo 1990 for a review). The $\delta$ Sct stars
show similar characteristics to the SX Phe stars but have solar
metallicities and more likely to be located in the thin disk (see
Breger 2000 for a review).  Although $\delta$ Sct-like pulsations have
been detected in pre-main sequence stars with periods as short as 18
min (Amado et al 2004), approximately 90 percent of $\delta$ Sct stars
have a pulsation period in the range 40 min to 5.3 hrs (cf Table 1,
Rodr\'{i}guez et al 2000).  SX Phe and $\delta$ Scu stars have a well
defined period-luminosity relationship and can therefore be used as
distance indicators and hence map Galactic structure (eg Nemec,
Linnell Nemec \& Lutz 1993).

Since we took our first set of data in 2003, we have obtained a
significant amount of further data (Barclay et al 2011). We therefore
have made a systematic search for blue, short period, pulsating
stars. Our light curves are typically 2--2.5 hrs in duration, so the
longest period we can determine with confidence is less than 2 hrs.
For stars with periods shorter than 40 mins, it becomes increasingly
difficult to determine the nature of the source based on colour and
period information (cf Table 1 and 2, Barclay et al 2011). In this
paper, we therefore have decided to restrict our search for blue
pulsating variables in the range of 40 min to 2 hrs.

\section{Observations}

\begin{figure}
\begin{center}
\setlength{\unitlength}{1cm}
\begin{picture}(14,6.5)
\put(-0.8,-0.5){\includegraphics{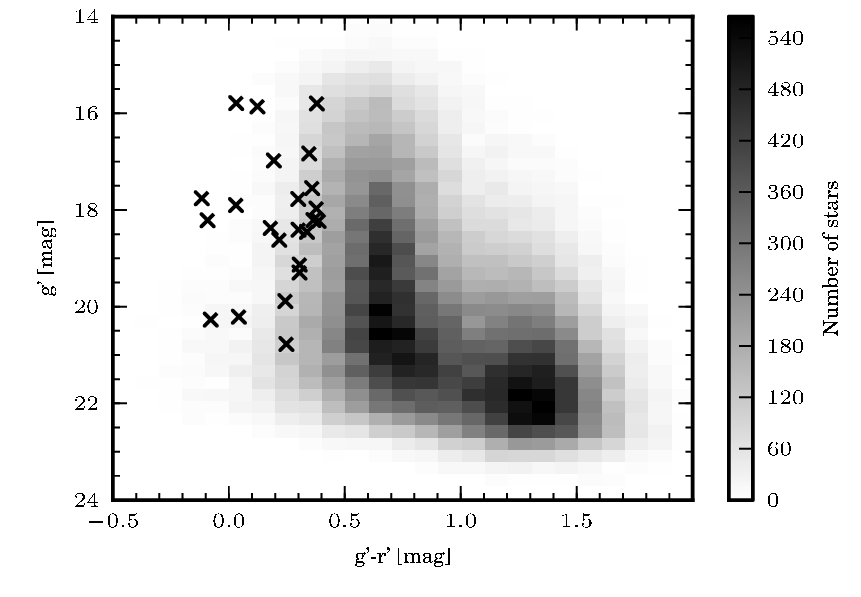}}
\end{picture}
\end{center}
\caption{The colour-magnitude diagram of all stars in the RATS archive
which are located in field with extinction less than $A_{V}\sim$0.4.
The blue pulsators presented in this paper are shown in Table
\ref{candidates} are shown as crosses.}
\label{colmag}
\end{figure}

The RATS observing strategy is to take a series of 30 sec exposures of
a given field using the wide-field cameras on the Isaac Newton
Telescope in La Palma and the MPG/ESO 2.2m on La Silla for a duration
of $\sim$2 hrs. To date our survey has discovered around
1.2$\times10^{5}$ variable stars (see Barclay et al 2011 for a full
description of our reduction process). Based on their photometric
properties, a small sub-sample has been selected for followup
spectroscopic observations to determine their nature.

To narrow our search for blue pulsators in our RATS data, we
restricted our search to a range in both magnitude and colour.  The
intrinsic colour of SX Phe stars is typically $(B-V)_{o}\sim$0.1 to
0.35 (eg Poretti et al 2008), which corresponds to $(g-r)_{o}=$--0.12
to 0.14. This is virtually identical to the colours for $\delta$ Sct
stars (eg Rodr\'{i}guez et al 2000). Many of our fields lie at low
Galactic latitude and hence have high extinction. To reduce the
contamination with other types of sources at low Galactic latitudes we
allow a maximum extinction corresponding to $A_{V}$=0.40
($E_{B-V}$=0.13 for R=3.1). For blue stars $E_{B-V}$=0.13 equates to
$E_{g-r}$=0.13. If we add a conservative uncertainty of 0.13 mag in
our observed colours (Barclay et al 2011) our search region therefore
covers $-0.12<(g-r)<0.40$, while the brightness of stars were in the
range $15<g<23$.

Here we note that the shape of the light curves and the colour of our
sources is similar to some cataclysmic variables (CVs; eg Szkody et al
2002 who present the first sample of CVs discovered using SDSS
data). The hydrogen accreting CVs have a minimum orbital period of
$\sim$80 mins (G\"{a}nsicke et al 2009), implying a possible over-lap
with our target selection. (The helium dominated CVs with orbital
periods in the range 40--70 mins do not show a photometric modulation
implying we are not overlapping with these objects).  However, CVs
show optical spectra dominated by line emission -- although we
obtained a spectrum for only 7 out of the 31 new sources -- none show
evidence for emission lines (cf \S \ref{spectra}). Moreover, although
pulsating blue stars are not X-ray sources, CVs are weak to moderately
strong X-ray sources (Verbunt et al 1997). We therefore
cross-correlated the position of our sources with that of catalogues
derived from the {\sl Rosat} All-Sky X-ray Survey (RASS). None of our
sources has an X-ray counterpart within 20 arcsec of the optical
position. In contrast 30 of the 48 CVs with orbital periods in the
range 70--120 mins in the 2009 catalogue of Ritter \& Kolb (2003) were
detected in the RASS. Although at this stage we cannot preclude that
none of our 31 sources are CVs we consider this rather unlikely.

\begin{table*}
\begin{center}
\begin{tabular}{lccrrcrrlrrlr}
\hline
Short &  $\alpha$  & $\delta$ &  $l$ & $b$ &  $g$  & $g-r$ & Period & Amp & $d$, $z$ & $d$, $z$ & Spec? & Low\\      
ID&  (J2000) &  (J2000) & (J2000) & (J2000) & & & (mins)  & (mag) & (kpc) & (kpc) & & $A_{V}$?\\
  &          &          &         &         & & &         &       & FM & FO & & \\
\hline
J000114&  00:01:14.7 & +53:43:05.4 &115.46 & -8.42 &17.8 &-0.12 &  64.3 & 0.50  &    6.0, -0.9 &      5.0,  -0.7&    &  \\
J000134&  00:01:34.8 & +53:51:42.2 &115.54 & -8.29 &17.0 & 0.19 &  73.5 & 0.07  &    4.2, -0.6 &      3.5,  -0.5&    &  \\
J000147&  00:01:47.5 & +53:23:18.7 &115.48 & -8.76 &16.8 & 0.35 &  69.6 & 0.45  &    3.4, -0.5 &      2.8,  -0.4&    &  \\
J030556&  03:05:56.8 &--00:36:16.2 &179.16 &-48.23 &20.3 &-0.08 &  82.9 & 0.28  &   30.0,-22.4 &     24.8, -18.5&SDSS& Y  \\
J050351&  05:03:51.4 & +35:08:02.1 &169.90 & -3.83 &18.4 & 0.30 &  62.9 & 0.14  &    3.4, -0.2 &      2.8,  -0.2&    &  \\
J065521&  06:55:21.2 & +10:41:58.9 &203.82 &  5.72 &15.9 & 0.12 &  60.9 & 0.13  &    2.7,  0.3 &      2.2,   0.2&    &  \\
J065541&  06:55:41.8 & +10:44:21.5 &203.82 &  5.81 &19.3 & 0.30 &  56.0 & 0.10  &   11.6,  1.2 &      9.6,   1.0&    &  \\
J120232&  12:02:32.2 &--24:29:17.4 &288.96 & 37.05 &20.2 & 0.04 &  55.7 & 0.33  &   21.2, 12.8 &     17.5,  10.6&    &  Y \\
J120709&  12:07:09.2 &--22:54:49.5 &289.81 & 38.82 &17.8 & 0.30 &  74.7 & 0.24  &    8.2,  5.2 &      6.8,   4.3&SAAO&  Y \\
J120902&  12:09:02.4 &--23:11:39.0 &290.43 & 38.65 &17.2 & 0.39 &  72.3 & 0.17  &    5.8,  3.6 &      4.8,   3.0&    &  Y \\
J135646&  13:56:46.3 & +22:54:40.4 & 20.61 & 74.62 &18.2 &-0.09 &  53.8 & 0.13  &    9.0, 8.7 &       7.4,  7.2&NOT &  Y \\
J135912&  13:59:12.9 & +23:36:55.3 & 23.76 & 74.30 &17.9 & 0.03 &  63.9 & 0.07  &    8.7,  8.3 &      7.2,   6.9&NOT &  Y \\
J155955&  15:59:55.4 &--25:43:20.2 &347.69 & 20.35 &17.3 & 0.37 &  60.8 & 0.08  &    5.2,  1.8 &      4.3,   1.5&    &  Y \\
J160103&  16:01:03.5 &--25:42:44.6 &347.89 & 20.17 &17.6 & 0.36 &  58.7 & 0.08  &    5.7,  2.0 &      4.7,   1.6&SAAO&  Y \\
J175816&  17:58:16.2 & +28:17:52.8 & 53.97 & 23.35 &19.1 & 0.04 &  54.4 & 0.12  &   12.7,  5.0 &     10.5,   4.1&WHT &  Y \\
J175836&  17:58:36.5 & +28:09:13.0 & 53.85 & 23.23 &18.1 & 0.05 &  66.1 & 0.07  &    9.2,  3.6 &      7.6,   3.0&WHT &  Y \\
J175932&  17:59:32.1 & +01:19:40.4 & 28.14 & 12.14 &18.2 & 0.36 &  74.7 & 0.50  &    6.5,  1.4 &      5.4,   1.1&    &  \\
J180331&  18:03:31.0 & +02:08:40.2 & 29.34 & 11.63 &17.8 &-0.09 &  72.5 & 0.65  &    7.0,  1.4 &      5.8,   1.2&    &  \\
J180416&  18:04:16.2 & +02:08:32.4 & 29.43 & 11.47 &18.1 & 0.23 &  82.9 & 0.65  &    8.2,  1.6 &      6.8,   1.4&    &  \\
J181727&  18:17:27.9 & +06:34:01.0 & 34.97 & 10.53 &18.3 & 0.22 &  76.9 & 0.10  &    8.6,  1.6 &      7.1,   1.3&    &  \\
J181736&  18:17:36.2 & +06:24:26.0 & 34.84 & 10.42 &18.6 & 0.22 &  60.8 & 0.06  &    8.2,  1.5 &      6.8,   1.2&    &  \\
J181753&  18:17:53.2 & +06:31:49.5 & 34.98 & 10.41 &18.5 & 0.34 &  59.6 & 0.08  &    7.4,  1.3 &      6.1,   1.1&    &  \\
J181816&  18:18:16.0 & +07:30:43.0 & 35.92 & 10.77 &18.5 & 0.24 &  53.0 & 0.50  &    7.1,  1.3 &      5.9,   1.1&    &  \\
J182250&  18:22:50.6 & +07:54:36.8 & 36.79 &  9.93 &19.1 & 0.31 &  64.5 & 0.13  &   10.8,  1.9 &      8.9,   1.5&    &  \\
J182347&  18:23:47.7 & +07:53:45.5 & 36.88 &  9.71 &19.9 & 0.24 &  61.2 & 0.26  &   15.7,  2.6 &     13.0,   2.2&    &  \\
J195235&  19:52:35.3 & +18:43:54.8 & 56.56 & -4.34 &18.0 & 0.38 &  63.2 & 0.05  &    4.9, -0.4 &      4.1,  -0.3&    &  \\
J200210&  20:02:10.0 & +18:43:07.3 & 57.72 & -6.29 &18.2 & 0.39 &  74.5 & 0.07  &    5.1, -0.6 &      4.2,  -0.5&    &  \\
J220915&  22:09:15.4 & +55:34:38.5 &101.37 & -0.35 &15.8 & 0.03 &  51.6 & 0.07  &    0.6,  0.0 &      0.5,   0.0&    &  \\
J230507&  23:05:07.9 & +34:17:23.4 & 99.14 &-23.62 &20.8 & 0.25 &  75.3 & 0.21  &   32.1,-12.8 &     26.5, -10.6&    &  Y \\
J233907&  23:39:07.3 & +57:08:02.3 &113.21 & -4.36 &18.4 & 0.18 &  67.6 & 0.15  &    4.1, -0.3 &      3.4,  -0.3&    &  \\
J234635&  23:46:35.6 & +56:23:23.7 &114.00 & -5.35 &15.8 & 0.38 &  68.6 & 0.18  &    1.8, -0.2 &      1.5,  -0.1&    &  \\
\hline
\end{tabular}
\end{center}
\caption{Candidate blue pulsating stars identified in RATS data. We
  show: the stars `short' ID; equatorial and galactic co-ordinates; $g$
  magnitude and $g-r$ colour; period and full-amplitude of
  modulation. We also show the distance from the Sun, $d$, and
    the height above the Galactic plane, $z$, (where FM implies we
    assume the period we detect is the fundamental period and FO
    imples the period is the first overtone) and whether we have a
  spectrum and if so its origin (\S 5). The last column indicates
  whether it is located in a field with low extinction -- they have
  been used in estimating the Galactic population of SX Phe stars (\S
  7).}
\label{candidates}
\end{table*}

For sources which fell within our search range we manually inspected
each light curve to verify that the light curve was consistent with
that of a stellar pulsator (cf Rodr\'{i}guez et al 2007 for a recent
example of a light curve for an SX Phe star) and to exclude light
curves of low quality. We found 31 sources which showed a modulation
in their lightcurve on a period between 51--83 mins and had a mean
brightness of $g=15.9-20.8$. The full-amplitude of the modulation is
in the range 0.05--0.65. We show their photometric properties in Table
\ref{candidates} and the light curves in Figure \ref{curves1} and
\ref{curves2}.

\section{Distances}
\label{distance}

There is a clear relationship between $M_{V}$ and pulsation period
which is applicable to SX Phe, $\delta$ Sct and RR Lyr stars (eg
McNamara 1997). This Period-Luminosity (PL) relationship is consistent
with a study of different types of short period pulsators in the
Fornax dwarf spheroidal galaxy (Poretti et al 2008). Since the PL
relationship of McNamara (1997) is calibrated with respect to $M_{V}$,
we applied a small correction to transform our $g$ band magnitudes to
that of the $V$ band (Jester et al 2005). Further, we used a NASA/IPAC
tool\footnote{http://irsa.ipac.caltech.edu/applications/DUST} which
uses the maps of Schlegel, Finkbeiner \& Davis (1998) to determine
extinction to the edge of the Galaxy.

The PL relationship assumes that the period is the fundamental radial
pulsation mode rather than the first over-tone which can also be
observed in these stars. Given the short duration of our light curves,
it is difficult to assess whether the period we detect is either the
fundamental or first over-tone period (or whether the period is even
due to a radial mode).  Some help is found from the fact that the
period of the first over-tone is less than the period of the
fundamental period by a factor of 0.775 (Poretti et al 2005).  We
determined the distance to each source assuming the period we detect
was the fundamental radial mode and also by assuming the period was
the first over-tone (we show the distance and corresponding
height from the Galactic plane to each source in Table
\ref{candidates} under both assumptions). The error on the distance
assuming we do not know the pulsation mode of the star is $\sim$17
percent.

\begin{figure}
\begin{center}
\setlength{\unitlength}{1cm}
\begin{picture}(8,6)
\put(-0.8,-0.4){\includegraphics{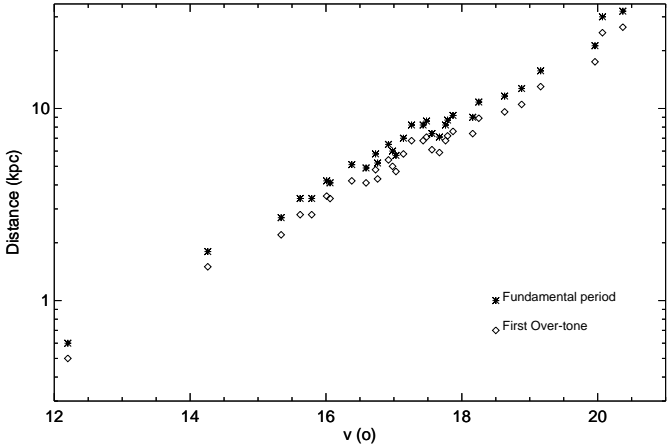}}
\end{picture}
\end{center}
\caption{The relationship between the dereddended $V$ mag and the 
distance determined assuming the measured period is the fundamental 
mode or the first over-tone. It suggests our distance determinations 
are not grossly in error.}
\label{dist-vo}
\end{figure}

\begin{figure*}
\begin{center}
\setlength{\unitlength}{1cm}
\begin{picture}(14,19)
\put(-0.4,-1.5){\includegraphics{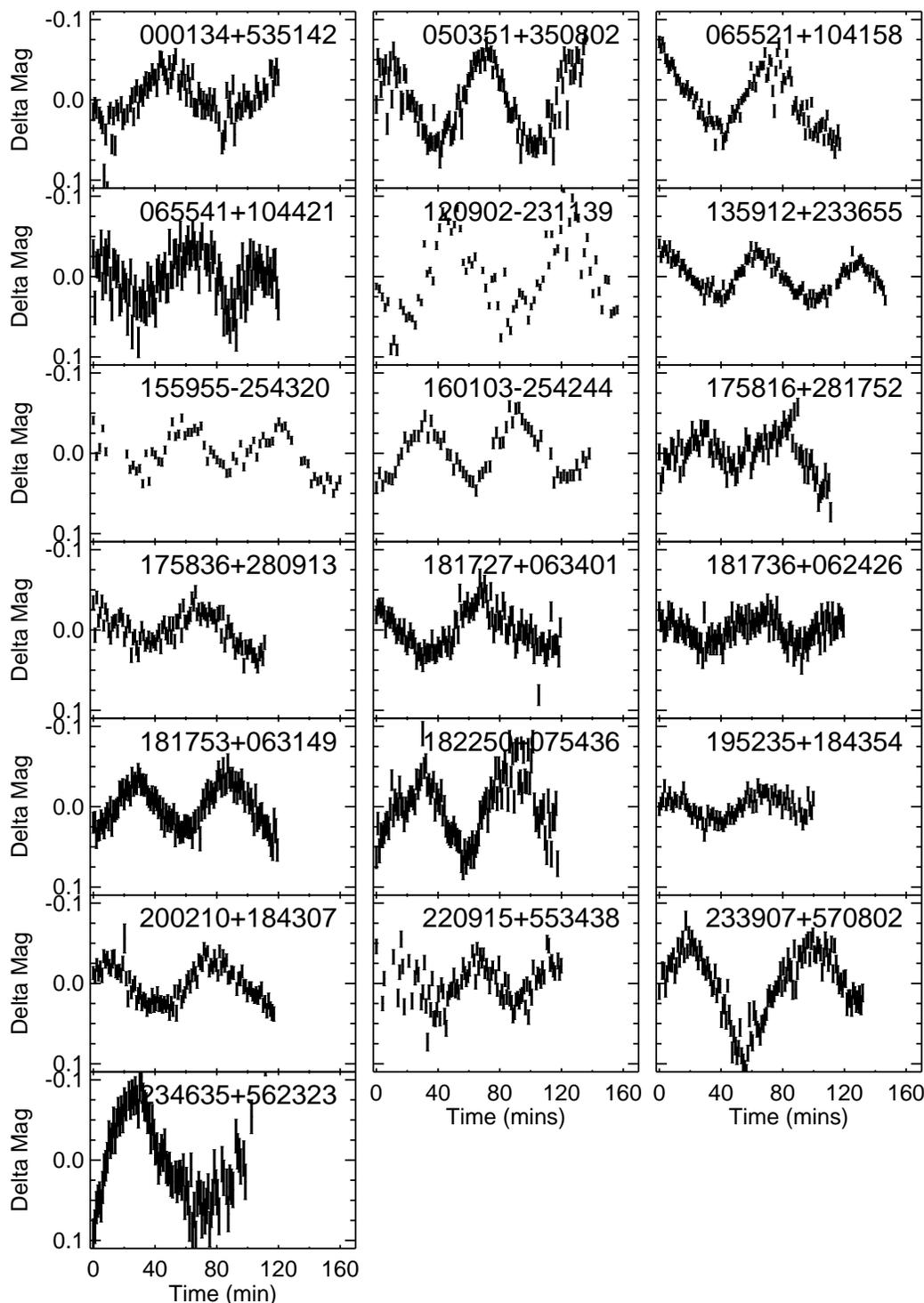}}
\end{picture}
\end{center}
\caption{The light curves of our blue candidate stellar pulsators
identified in our survey and which had an amplitude less than 0.2
mag.}
\label{curves1}
\end{figure*}

\begin{figure*}
\begin{center}
\setlength{\unitlength}{1cm}
\begin{picture}(14,12)
\put(-0.4,-9){\includegraphics{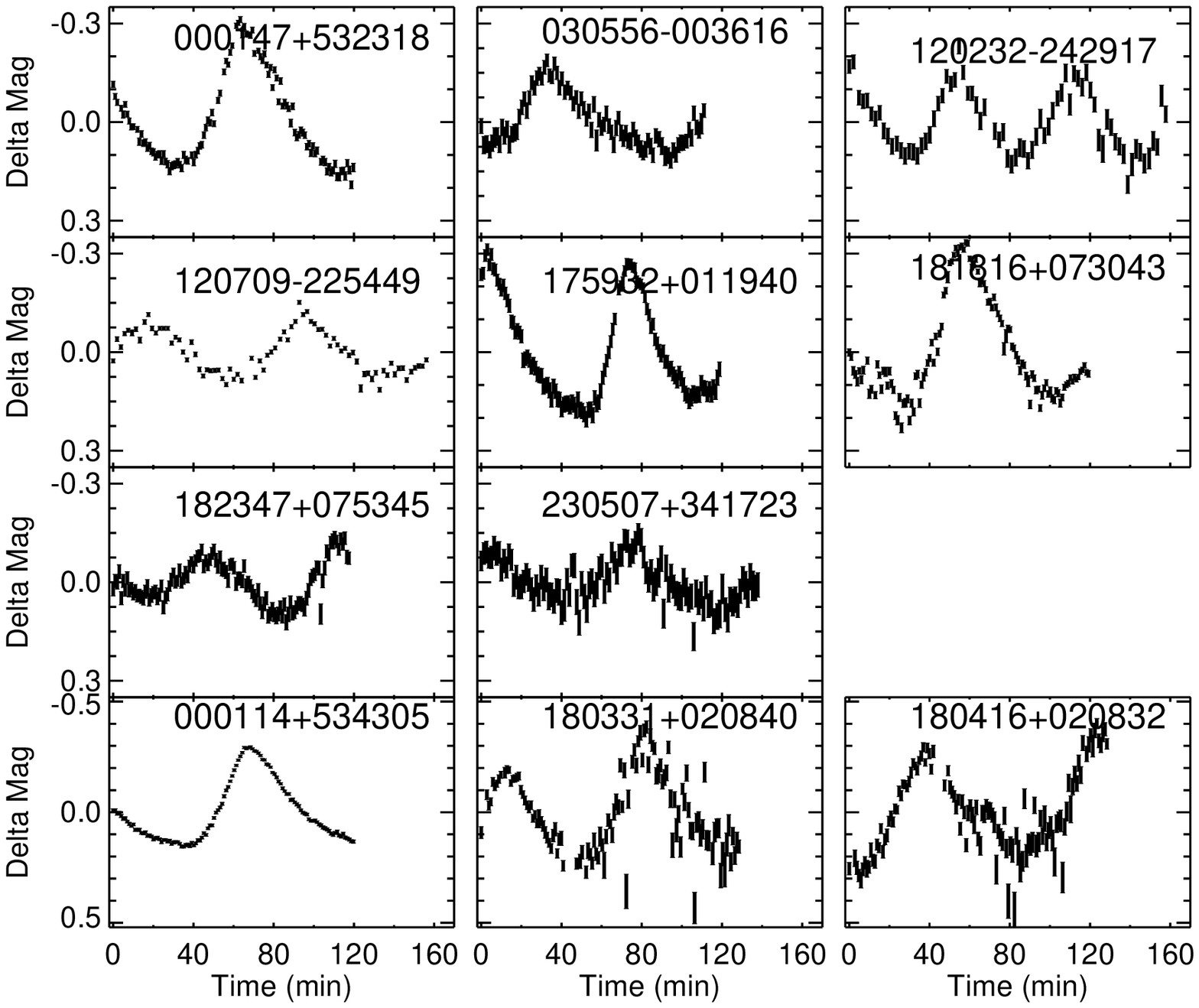}}
\end{picture}
\end{center}
\caption{As for Fig \ref{curves1} but for those stars which had an
amplitude greater than 0.2 mag.}
\label{curves2}
\end{figure*}

As a zeroth order test, we show in Figure \ref{dist-vo} the
relationship between the dereddened $V$ mag (assuming the extinction
to the edge of the Galaxy as determined above) and the derived
distance assuming the period is the fundamental mode and also the
first over-tone. Whilst one can argue that any individual object may
give a better overall linear relationship if one assumes the period is
one or the other mode, it gives us confidence that our distances are
not grossly in error.

Taking into account the uncertainties in our photometry ($g\sim0.1$
for stars brighter than $g$=20, increasing to $g\sim0.2$ for fainter
sources), the uncertainty on our period determinations, coupled with
the uncertainty on the pulsation mode, we estimate the errors on our
distances maybe up to 25 percent.  If on the other hand the periods we
determine are half the true period then we significantly underestimate
their distances. Similarly, if the extinction is less than that to the
edge of the Galaxy then we also underestimate the distances. Of
course, if the period we detect is not due to radial pulsation then
the distance is highly uncertain.

Our sources have a large spread of distances (Table 1). Assuming the
period is due to the fundamental radial pulsation mode then the
closest star is 0.6 kpc, while the most distant lies at 32 kpc (the
median is 7.0 kpc).  Similarly, the sample shows a large spread in
height from the Galactic plane, the least distant only 200 pc, with
the most distant at 20 kpc. The median height is 1.4 kpc, which
equates to twice the scale height of the thick disc of the Galaxy (eg
de Jong et al 2010). On the other hand, if the periods are due to 
the first over-tone radial mode then these distances are less by 
$\sim$20 percent.

The two sources with the greatest distances (J030556 and J230507 at
$\sim$30 kpc) have a Galactic longitude which appears to place them at
a distance much further than the accepted limits of the spiral
structure of the Milky Way (eg Churchwell et al 2009). Secondly, three
sources (J030556, J120232, J230507) lie more than 10 kpc distant from
the Galactic plane: this places them deep into the Galactic halo.

\section{Followup Photometry}

\begin{figure}
\begin{center}
\setlength{\unitlength}{1cm}
\begin{picture}(8,6)
\put(-0.5,0){\includegraphics{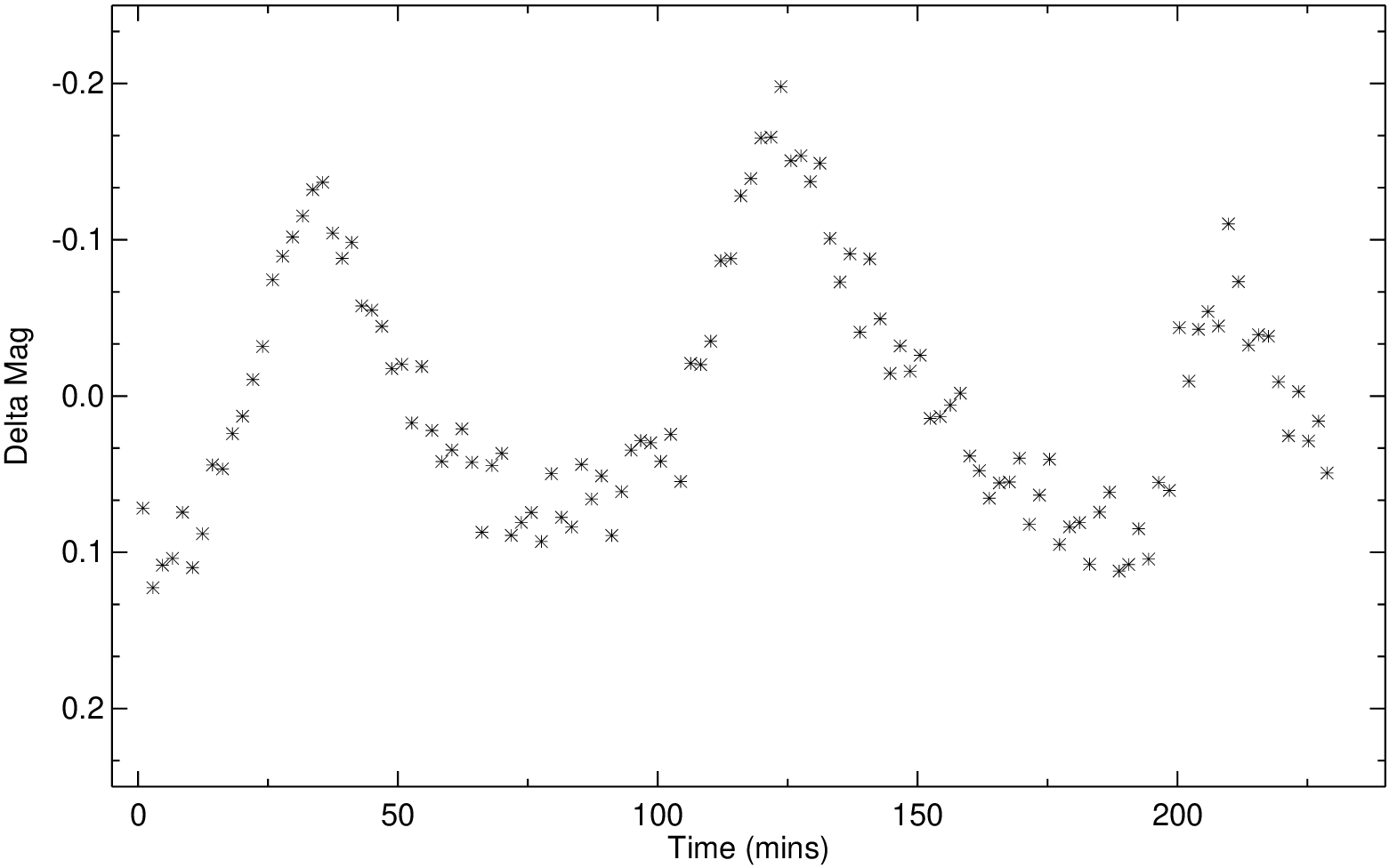}}
\end{picture}
\end{center}
\caption{The light curve of RAT J030556-003616 made using the NOT and
ALFOSC in Dec 2010. The data has been binned into 120 sec bins.}
\label{j0305}
\end{figure}

To confirm the period of J0305, we obtained followup photometry of
this source on 31 Dec 2010 using the Nordic Optical Telescope and
ALFOSC. We used white light and an exposure time of 10 sec. The
resulting light curve, binned into 120 sec bins, covers 3.8 hrs
(Figure \ref{j0305}). A clear modulation is present in the light
curve. A Lomb Scargle power spectrum of the light curve indicates a
period of 90 min.  This compares with 83 min (Table 1) which was
derived using the original INT light curve. Given the uncertainties in
the period derived using each data-set, the periods are
consistent. For completeness, we note that using a period of 90 min
rather than 82.9 min (Table 1) places J0305 at a distance of 31.9 kpc
rather than 30.0 kpc assuming we have identified this period as
  the fundamental radial pulsation mode. We encourage additional
  photometry of all the sources shown in Table 1 to identify the mode
  of the pulsation seen in each star.

\section{Spectral observations}
\label{spectra}

We have spectroscopic observations of a small sample of our candidate
blue pulsating stars (cf Table 1). We obtained spectra for two sources
using the 4.2m William Herschel Telescope (WHT) and the Intermediate
dispersion Spectrograph and Imaging System (ISIS) on La Palma, two
using the SAAO 1.9m telescope and the Cassegrain spectrograph in
Sutherland, South Africa and two using the 2.5m Nordic Optical
Telescope and ALFOSC on La Palma. A spectrum of another source was
obtained from the SDSS data archive\footnote{http://www.sdss3.org/dr8}.

Both arms of ISIS were used giving spectral coverage from
$\sim$3800--5200 \AA\ and $\sim$5500--9000 \AA\ in the blue (R158B
grating) and red (R158R) arms respectively. Grating \#7 was used with
the SAAO spectrograph giving a wavelength range of
3400--7500\AA. The spectral resolution was $\sim$5\AA\hspace{1mm}
for the WHT and SAAO spectra. Grating \#7 was used with the ALFOSC
imaging spectrograph giving a wavelength range
3800--7000\AA\hspace{1mm} and a spectral resolution of
$\sim$8\AA. All the data were reduced using optimal extraction and
standard techniques.  Several spectra of J0305 were downloaded from
the SDSS archive and co-added and re-binned into 4 \AA\ bins.

We modelled the spectra using a grid of LTE models calculated with the
{\sc atlas9} code (Kurucz 1992) with convective overshooting switched
off. Spectra were calculated with the {\sc linfor} line-formation code
(Lemke 1991). Data for atomic and molecular transitions were compiled
from the Kurucz line list. The spectra were fitted with the {\sc
fitsb2} routine (Napiwotzki et al 2004). The error limits of all fit
parameters were determined with a bootstrapping method.

The stellar temperatures were estimated from the hydrogen Balmer lines
of the stars (H$\beta$ to H$\epsilon$). No gravity sensitive features
are accessible in our low resolution spectra so gravity was fixed at
$\log g = 4.0$, which is a typical value for SX Phe and large
amplitude $\delta$ Sct stars (eg McNamara 1997). 

Apart from the Ca H and K lines, the resolution of our spectra is
clearly too low to allow a meaningful fitting of individual metal
lines for abundance determinations. However, the spectra allow a
determination of an overall abundance value. The general metallicity
of the models was varied until an optimum fit was achieved. Since the
extinction towards these high Galactic latitude targets is low we do
not expect the Ca H and K lines to be significantly contaminated by
interstellar absorption. However, by including these lines we obtain
an upper limit to the metallicity.  The spectral ranges used for
determining the metallicity contain a mix of spectral features, but
the dominant species is Fe I. We thus expect our metallicity [Met/H]
to be an approximate indicator of iron abundance.

\begin{figure*}
\begin{center}
\setlength{\unitlength}{1cm}
\begin{picture}(14,10)
\put(-3,0){\includegraphics{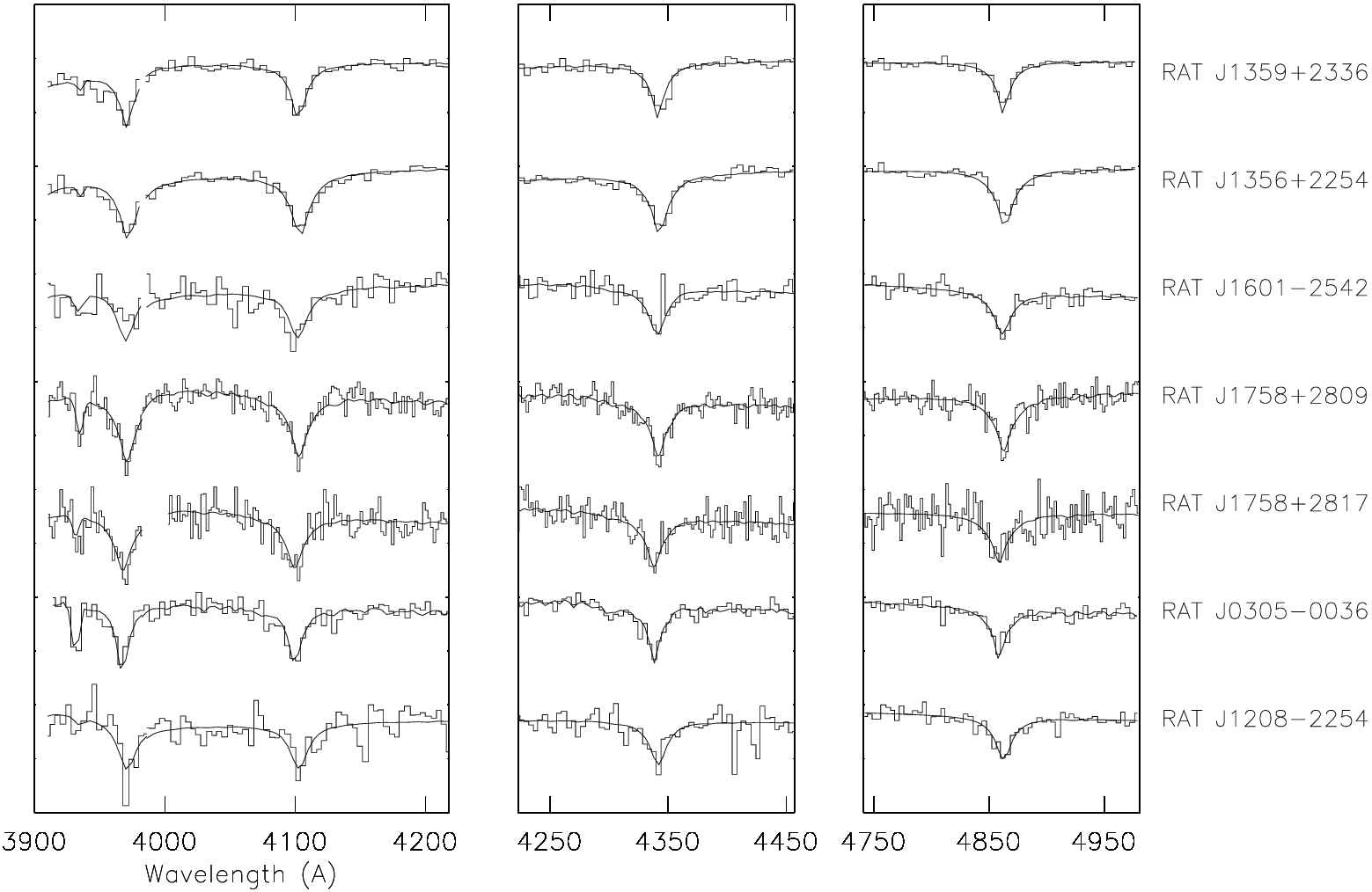}}
\end{picture}
\end{center}
\caption{We show the spectra and best model fit for those sources for which
spectral data was available.}
\label{spec-fit}
\end{figure*}

We show the results of our fits in Table~2 and the best fits to the
spectra in Figure \ref{spec-fit}. The best-fit temperature of our
sources are in the range \teff=7200--7900K. Although the metal
abundance is less well constrained, each source has a metallicity less
than solar at the 1$\sigma$ level and for all but one source this is
also true at the 3$\sigma$ level. Fixing $\log g$ at 3.5 and 4.5
rather than 4.0 changes the temperature by less than 90K and $\log g$
by less than 0.1 dex.

\begin{table}
\begin{center}
\begin{tabular}{lrr}
\hline
Source &  Temp (K) & Z (\Msun) \\
\hline
J030556 &  7210$^{+270}_{-230}$ & -0.5$^{+0.4}_{-2.3}$  \\
J120709 &  7310$^{+480}_{-350}$ & -2.1$^{+1.8}_{-3.3}$ \\
J135646 &  7750$^{+270}_{-240}$ & -1.1$^{+0.6}_{-3.5}$ \\
J135912 &  7210$^{+200}_{-290}$ & -2.1$^{+1.0}_{-2.1}$ \\
J160103 &  7410$^{+310}_{-360}$ & -1.1$^{+1.1}_{-4.0}$ \\
J175816 &  7830$^{+680}_{-400}$ & -1.0$^{+0.9}_{-4.7}$ \\
J175836 &  7890$^{+280}_{-230}$ & -1.3$^{+1.1}_{-3.6}$  \\
\hline
\end{tabular}
\end{center}
\caption{The temperature and metallicity for seven of our sources
derived from model fits to their optical spectra. The errors refer to the
3$\sigma$ confidence interval.}
\end{table}

\section{The nature of the variable sources}
\label{nature}

We have presented evidence that the majority of sources described in
this paper are blue compact stellar pulsators. To place our sources in
the general context of blue stellar pulsators, we used the most recent
edition of the General Catalogue of Variable Stars (GCVS, Samus et al
2009) to determine the distribution of periods in different classes of
short period pulsating variables.  SX Phe and $\delta$ Sct stars and
the longer period pulsating sdB stars all have a distribution of
pulsation periods which include periods less than 60 mins. In
contrast, RR Lyr and classical Cepheids have much longer pulsation
periods.  The $\beta$ Cepheid stars have a minimum period of $\sim$2
hrs.  The GCVS notes 131 field $\delta$ Sct stars and 20 SX Phe stars.

SX Phe stars and $\delta$ Sct stars have temperatures typically
7200--7900K (implying late A/early F spectral types), while the long
period sdB stars have temperatures typically $\sim$25000--30000K
(Green et al 2003). Our analysis of those sources for which we have
spectra result in parameters which confirm a SX Phe/$\delta$ Scuti
nature (\S \ref{spectra}). The observed colours are also consistent
with this classification for all pulsators, although for fields with
high extinction we cannot rule out some contamination from
intrinsically redder sources. However, given that the amplitude of the
long period sdB stars is very low (eg Fontaine et al 2003) we consider
it highly unlikely that blue pulsators are sdB stars.

The light curves of our sources (Figure 2 and 3) appear to show
regular pulsations periods, some displaying high ($\sim$60 percent)
full-amplitude modulations, while others are much lower (a few
percent). They are similar to the light curves of SX Phe and $\delta$
Sct stars which appear in the literature. We have low resolution
spectra for a small number of our sources which were taken for
identification purposes: they are consistent with A/F type
stars. Although our spectra are not high resolution, our model fits
indicate that all sources have metallicities which are less than Solar
(for most sources at the 3$\sigma$ confidence level). Higher
resolution spectra with good signal-to-noise are necessary to
determine their metal content with higher confidence.

In section \ref{distance} we noted that the vast majority of our
sources are located at distances greater than 3 kpc, with some being
over 20 kpc distant (if we have detected either the fundamental
  or first over-tone mode). Similarly, many are at a height of 1 kpc
or more from the Galactic plane, with several being at least 10 kpc
from the plane. Some of our sources are therefore at the remote edge
of our Galaxy. SX Phe stars have been found at both large distances
and well into the Galactic plane (eg Berstein, Knezek \& Offutt 1995;
Jeon, Kim and Nemec 2010). They are therefore, in principal, potential
tracers of Galactic structure such as streams or the remnants of
mergers.

We note that J1356 and J1359 lie around $\sim6^{\circ}$ from the large
globular cluster M3 (which is 10.4 kpc from the Sun) which places them
at a distance of $\sim$2 kpc from M3. Although tidal tails have been
detected at distances of several kpc from globular clusters (eg
Odenkirchen et al 2003), no tidal tails have been detected from M3
(Jordi \& Grebel 2010). The other source of particular note is J0305
which is located in the direction of the Galactic anti-center and at a
distance of 30kpc (implying a Galactocentric distance of 38 kpc) and a
height of 21 kpc below the Galactic plane. Recent work shows that the
stellar density of the Galaxy decreases sharply at Galactocentric
distances greater than 25 kpc (eg Watkins et al 2009, Sesar et al,
2010) which may indicate that J0305 is associated with a sub-structure
of the halo. Alternatively it may be in the process of being ejected
from our Galaxy. A more detailed radial velocity study is required to
answer this question.

\section{The Galactic population of SX Phe stars}

SX Phe stars have been identified in globular clusters and nearby
dwarf galaxies (eg Olech et al 2005, Poretti et al 2008). Based on the
number of blue stellar pulsators which we have identified in our
survey, we now make an estimate of the total number of SX Phe stars
which are present in our Galaxy. Since our survey is biased towards
low Galactic fields, highly reddened (but intrinsically blue)
pulsators could be confused with apparently much redder sources making
potential contamination a significant concern. For this reason, we
therefore base our simulation on those fields for which the total
extinction is less than $A_{V}$=0.45. A total of 35 fields had a
column density less than our limit which corresponds to an area of 10
square degrees. We find eleven blue pulsators in these fields (these
are flagged in the last column of Table 1).

Low resolution spectroscopic data exist for seven of these eleven
pulsators and a spectroscopic analysis indicates they have a metal
content consistent with that of SX Phe stars. Here we assume that all
eleven pulsators at high Galactic latitudes are SX Phe stars.  We use
a simulation of the Galactic populations found in the fields to
extrapolate the number of SX Phe stars found in the whole Galaxy.  Our
simulation uses a modified version of a model originally developed for
populations of hot evolved stars (Napiwotzki 2008). Thin disc, thick
disc and halo populations are included in our simulations, while the
scale height adopted for the thick disk is 800 pc. The halo is
modelled as an oblate ellipsoid with axis ratio $\epsilon$ =
0.76. Reddening is included in a simple approximation. Further details
are given in Table~3 of Robin et al.\ (2003).  No self-consistant
modelling of the evolution of main sequence stars and possible binary
channels, which might lead to the formation of SX~Phe stars, is
performed. We assume that the absolute brightness of SX~Phe is
randomly distributed within the observed interval $M_V=2.1-3.2$ (which
is the implied absolute magnitude for SX Phe stars with periods of
2--1 hr respectively, McNamara 1997) and that the space density of
SX~Phe is simply proportional to the density of the parent
population. This should be a good approximation for the field
population in which dynamical interaction plays a very small role.

SX~Phe stars are observed in globular clusters and are known to be
metal poor. Thus it is clear that the thin disk can be ruled out as
the parent population, but both the thick disk and halo populations,
are feasible. The thick disk population is almost as old as the
Galactic halo and some stars of this population have quite low
metallicities. However, some thick disk stars have overall abundances
comparable to thin disk stars (Bensby, Feltzing \& Lundstr\"{o}m 2003,
Fuhrmann 2004).  Depending on the degree of this fraction and noting
that the SX~Phe phenomenon is linked to metallicity, different
formation efficiencies can be expected. We make two extreme
assumptions to constrain the Galactic SX~Phe population: 1) thick disk
and halo have the same formation efficiency or 2) SX~Phe stars are
only formed from halo stars.

A total of 250 million SX~Phe was simulated, making the statistical
error of the Monte Carlo simulation negligible. We obtained a
cumulative distribution of SX Phe stars as function of limiting
magnitude.  A catalogue of simulated stars in $1^\circ$ fields around
the central coordinates of the RATS fields and brighter than the
detection limit of SX~Phe variables ($V=22$ [for blue stars this
  implies $g$=22]) was produced. Stars in this list were weighted with
the effective field of view of the cameras (Barclay et al
2011). Simulated star numbers were scaled according to the predicted
number of stars and the observed number of stars (11 in the low
extinction fields). The result is that we predict 6.6$\times10^{4}$
SX~Phe stars brighter than ($V=22$) in our Galaxy if a mix of halo and
thick disk is assumed and 4.0$\times10^{4}$ if only the halo
population contributes. 

Recent determinations of the dynamical mass of the Milky Way include
$\sim1\times10^{12}$ \Msun (Watkins, Evans \& An 2010) and
$\sim2.5\times10^{12}$ \Msun (Sakamoto et al 2003). However, given
that the mass of the Milky Way is thought to be dominated by dark
matter, the stellar mass is expected to be 1/20 of the dynamical mass
(eg Moore et al 1999), giving a stellar mass in the range
$\sim5-12\times10^{11}$ \Msun. Our simulations therefore imply one SX
Phe star per 7.6$-18\times10^{5}$ stars and that SX Phe stars are
significantly less abundant per unit mass in our Galaxy compared to
than that found in globular clusters (eg one per $\sim4\times10^{4}$
\Msun for $\omega$ Cen, Olech et al 2005) and dwarf spheroidal
galaxies (eg one per $2.1\times10^{4}$ \Msun for the Fornax dSph,
Poretti et al 2008). (We have assumed no dark matter is present in
globular clusters and used the results of $\L$okas 2009 in determining
the stellar mass of the Fornax dSph). Given we have assumed that all
eleven stars at low Galactic latitude are SX Phe stars (and the
spectra of seven are consistent with this) our estimate of the number
of SX Phe stars in our Galaxy may be an overestimate, indicating an
even greater discrepancy between the relative number of SX Phe stars
in our the Galaxy and other nearby stellar groups.

Although there is some uncertainty in the number of bona fide SX Phe
stars in our survey, the discrepancy between the numbers of SX Phe
stars predicted in our Galaxy and nearby stellar systems is over an
order of magnitude. The fact that SX Phe stars are less abundant in
our Galaxy is presumably a consequence of the metallicity and star
formation history of these systems. A comprehensive study of the
number of SX Phe stars in different environments could lead to a
better understanding of how these stars are formed.

\section{Conclusions}

We have identified 31 blue pulsating objects for which we have
evidence that they are candidate SX Phe or $\delta$ Sct stars. These
pulsators which have periods between 51--83 mins are well suited to
being discovered using surveys like RATS which have high cadence but
have a relatively short overall duration. Unlike the RR Lyrae stars
which have a longer pulsation period and corresponding brighter
absolute magnitude, they have been little used to identify Galactic
sub-structure. Our results suggest that existing survey telescopes
would be well suited to the discovery of SX Phe and $\delta$ Sct stars
if their cadence was high enough. Further, if the mode of pulsation
can be identified then they would provide a useful cross-calibration
set for luminosity-period relationships and how this is affected by
metallicity.

\section{Acknowledgements}

This paper is based on observations obtained using the Isaac Newton
Telescope, the William Herschel Telescope on La Palma (the ING); the
MPG/ESO 2.2m and ESO 3.6m telescopes at the European Southern
Observatory, La Silla, Chile under programmes 075.D-0111(A) and
079.D-0621(A). We thank the staff of the ING and ESO for their help in
obtaining these observations. This paper also uses data obtained using
the 1.9m telescope of the South African Astronomical Observatory,
Sutherland, South Africa and the 2.5m Nordic Optical Telescope. We
thank the referee for pointing out the issues surrounding the
identification of the mode of pulsation and Michael Lemke for the grid
of models spectra computed for a `A quick project'.

{}


\begin{thebibliography}{}

\bibitem{}Amado, P. J., Moya, A., Su\'{a}rez, J. C., Mart\'{i}n-Ruiz, S.,
Garrido, R., Rodr\'{i}guez, E., Catala, C., Goupil, M. J, 2004, MNRAS,
352, L11 
\bibitem{}Barclay, T., Ramsay, G., Hakala, P., Napiwotzki, R., Nelemans, G.,
Potter, S., Todd, I., 2011,, MNRAS, 413, 2696
\bibitem{}Bensby, T, Feltzing, S., Lundstr\"{o}m, I., 2003, A\&A, 410, 527
\bibitem{}Breger, M., 2000, ASP Con Series, 210, 3
\bibitem{}Churchwell, E., et al, 2009, PASP, 121, 213
\bibitem{}de Jong, J. T. A., Yanny, B., Rix, H.-W., Dolphin, A. E., 
Martin, N. F., Beers, T. C., 2010, ApJ, 714, 663
\bibitem{}Fontaine, G., Brassard, P., Charpinet, S., Green, E. M., 
Chayer, P., Billeres, M., Randall, S. K., 2003, ApJ, 597, 518
\bibitem{}Fuhrmann, K., 2004, Astronomische Nachrichten 325, 3
\bibitem{}G\"{a}nsicke, B. T., et al, 2009, MNRAS, 397, 2170
\bibitem{}Green, E. M., et al, 2003, ApJ, 583, L31
\bibitem{}Kurtz, D. W., 2004, Solar Physics, 220, 123
\bibitem{}Kurucz, R. L., 1992, Rev Mex Astron Astro., 23, 181
\bibitem{}Lemke, M. 1991, Internal Report, Department of Astronomy, University
of Texas at Austin
\bibitem{}$\L$okas, E. L., 2009, MNRAS, 394, L102
\bibitem{}Jeffery, C. S., 2008, CoAst, 157, 240
\bibitem{}Jeon, Y.-B., Lee, S.-L., Nemec, J. M., 2010, PASP, 122, 17  
\bibitem{}Jester, S., et al, 2005, AJ, 130, 873
\bibitem{}Jordi, K., Grebel, E. K., 2010, A\&A, 522, 71
\bibitem{}McNamara, D. H., 1997, PASP, 109, 1221
\bibitem{}Moore, B., Ghigna, S., Governato, F., Lake, G., Quinn, T., Stadel, J., 
Tozzi, P., 1999, ApJ, 524, L19
\bibitem{}Napiwotzki, R., et al, 2004, In {\sl Spectroscopically and
  Spatially Resolving the Components of the Close Binary Stars}, ASP Conf Series,
318, 402
\bibitem{}Napiwotzki, R., 2008, In {\sl Hot subdwarf stars and related objects}, 
ASP Conf Series, 392, 139 
\bibitem{}Nemec, J., Mateo, M., 1990, In {\sl Confrontation between
stellar pulsation and evolution}, ASP Con Series, 11, 64
\bibitem{}Nemec, J. M., Linnell Nemec, A. F., Lutz, T. E., 1993, ASPC, 53, 145
\bibitem{}Odenkirchen, M., et al, 2003, AJ, 126, 2385
\bibitem{}Olech, A., Dziembowski, W. A., Pamyatnykh, A. A., Kaluzny, J., 
Pych, W., Schwarzenberg-Czerny, A., Thompson, I. B., 2005, MNRAS, 363, 40
\bibitem{}Poretti, E., et al, 2005, A\&A, 440, 1097
\bibitem{}Poretti, E., Clementini, G., Held, E. V., Greco, C., Mateo, M., 
Dell'Arciprete, L., Rizzi, L., Gullieuszik, M., Maio, M., 2008,
ApJ, 685, 947
\bibitem{}Ramsay, G., Hakala, P., 2005, MNRAS, 360, 314
\bibitem{}Ramsay, G., Napiwotzki, R., Hakala, P., Lehto, H., 2006,
MNRAS, 371, 957
\bibitem{}Ritter, H., Kolb, U., 2003, A\&A, 404, 301
\bibitem{}Robin, A. C., Reyle, C., Derriere, S., Picaud, S., 2003, A\&A, 409, 
523
\bibitem{}Rodr\'{i}guez, E., L\'{o}pez-Gonz\'{a}lez, 
M. J., L\'{o}pez de Coca, P., 2000, A\&AS, 144, 469
\bibitem{}Rodr\'{i}guez, E. et al 2007, A\&A, 471, 255
\bibitem{}Samus, N. N., et al, 2009, adsabs.harvard.edu/abs/2009yCat....102025S
\bibitem{}Sakamoto, T., Chiba, M., Beers, T. C., 2003, A\&A, 397, 899
\bibitem{}Schlegel, D. J., Finkbeiner, D. P., Davis, M., 1998, ApJ, 500, 525
\bibitem{}Sesar, B., et al, 2010, ApJ, 708, 717
\bibitem{}Szkody, P., et al, 2002, ApJ, 123, 430
\bibitem{}Watkins, L. L., et al, 2009, MNRAS, 398, 1757
\bibitem{}Watkins, L. L., Evans, N. W., An, J. H., 2010, MNRAS, 406, 264
\bibitem{}Verbunt, F., Bunk, W. H., Ritter, H., Pfeffermann, E., 1997, A\&A,
327, 602
\end{thebibliography}
\end{document}